\documentclass[review]{elsarticle}

\usepackage{lineno,hyperref}

\usepackage{amsmath,amsthm,amsfonts,bm}	 
\usepackage{subfigure}
\usepackage{float}
\usepackage{multirow}

\journal{International Journal of Heat and Mass Transfer}

\begin{document}

\begin{frontmatter}

\title{Multi-GPU thermal lattice Boltzmann simulations using OpenACC and MPI
\footnote{\href{https://doi.org/10.1016/j.ijheatmasstransfer.2022.123649}{DOI: 10.1016/j.ijheatmasstransfer.2022.123649}}
\footnote{\copyright \ 2022. This manuscript version is made available under the CC-BY-NC-ND 4.0 license http://creativecommons.org/licenses/by-nc-nd/4.0/}}



\author[mymainaddress,mysecondaryaddress,mythirdaddress]{Ao Xu\corref{mycorrespondingauthor}}
\cortext[mycorrespondingauthor]{Corresponding author}
\ead{axu@nwpu.edu.cn}

\author[mymainaddress]{Bo-Tao Li}

\address[mymainaddress]{School of Aeronautics, Northwestern Polytechnical University, Xi’an 710072, China}
\address[mysecondaryaddress]{Institute of Extreme Mechanics, Northwestern Polytechnical University, Xi’an 710072, China}
\address[mythirdaddress]{Key Laboratory of Icing and Anti/De-icing, China Aerodynamics Research and Development Center, Mianyang 621000, China}

\begin{abstract}
We assess the performance of the hybrid Open Accelerator (OpenACC) and Message Passing Interface (MPI) approach for multi-graphics processing units (GPUs) accelerated thermal lattice Boltzmann (LB) simulation.
The OpenACC accelerates computation on a single GPU, and the MPI synchronizes the information between multiple GPUs.
With a single GPU, the two-dimension (2D) simulation achieved 1.93 billion lattice updates per second (GLUPS) with a grid number of $8193^{2}$, and the three-dimension (3D) simulation achieved 1.04 GLUPS with a grid number of $385^{3}$, which is more than 76\% of the theoretical maximum performance.
On multi-GPUs, we adopt block partitioning, overlapping communications with computations, and concurrent computation to optimize parallel efficiency.
We show that in the strong scaling test, using 16 GPUs, the 2D simulation achieved 30.42 GLUPS and the 3D simulation achieved 14.52 GLUPS.
In the weak scaling test, the parallel efficiency remains above 99\% up to 16 GPUs.
Our results demonstrated that, with improved data and task management, the hybrid OpenACC and MPI technique is promising for thermal LB simulation on multi-GPUs.
\end{abstract}


\end{frontmatter}


\section{Introduction}

Over the past half-century, the development of semiconductor transistors has driven rapid growth and prosperity of high-performance computing (HPC).
The miniaturization of computer components was foreseen by physicist Richard Feynman in his 1959 address "There is Plenty of Room at the Bottom" \cite{feynman1959plenty}.
Later in 1975, Gordon Moore, the founder of Intel Corporation, predicted that the number of transistors per computer chip would double every two years, which is known as Moore’s law \cite{moore1975progress}.
This trend held up considerably well until recently when transistors reduce their physical size and are reaching the atomic scale, implying that Moore’s law is nearing its end and is anticipated to flatten by 2025 \cite{shalf2020future}.
Alternative avenues for growth in computer performance include hardware architecture, software, and algorithms \cite{leiserson2020there}.
In the post-Moore era, computer architects should focus on hardware streamlining and provide additional chip area for more circuitry to operate in parallel, rather than use more transistors and increase the complexity of processing cores as they used to do.
For example, the graphics process unit (GPU) contains many parallel lanes and it can exploit much more parallelism, thus it delivers much more performance on computations.

The lattice Boltzmann (LB) method is a numerical approach to simulate fluid flows and associate heat and mass transfer processes \cite{chen1998lattice,aidun2010lattice}.
Specifically, the LB method describes the evolution of particle density distribution, which originated from the Boltzmann kinetic theory but practically reflects hydrodynamic behavior at the continuum scale.
Mesoscopic physical pictures can be easily incorporated into the LB method, and macroscopic physical conservation laws can be recovered with a relatively low computational cost.
After three decades of development of the LB method, the computational fluid dynamics community has witnessed its powerful ability to simulate complex flows, such as gas-liquid two-phase flow \cite{cheng2014recent,li2016lattice}, particulate flow \cite{maxey2017simulation,tao2022lattice,xiong2012large}, fluid-structure interaction \cite{xu2022free}, flow in porous media \cite{he2019lattice}, and so on.
Advancements in HPC utilizing heterogeneous architecture, namely the combined traditional central processing unit (CPU) and the emerging accelerators (such as GPUs), further facilitate the application of LB simulations in large-scale engineering problems \cite{liu2019sunwaylb,falcucci2021extreme}.
Open-source codes based on the LB method, including OpenLB \cite{krause2021openlb}, Palabos \cite{latt2021palabos}, and Sailfish \cite{januszewski2014sailfish}, even aim to brace the forthcoming Exascale supercomputing \cite{amati2021projecting}.
Review articles on LB simulation using GPUs can be found by Navarro-Hinojosa et al. \cite{navarro2018physically}, Niemeyer and Sung \cite{niemeyer2014recent}.

The pioneer works implementing LB simulations on graphics hardware were achieved by using textures and render buffers \cite{li2003implementing}.
Since 2007, with the introduction of Compute Unified Device Architecture (CUDA), which is a parallel computing platform and application programming interface (API), significant advances in applying LB simulation on GPUs have been made \cite{tolke2008teraflop,delbosc2014optimized,huang2015implementation,huang2015multi}.
Meanwhile, the Open Computing Language (OpenCL, initially released in 2009), which is an open standard for writing programs executing across the heterogeneous platform, has also accelerated the LB simulation without being restricted to only running on NVIDIA’s hardware.
For example, Sailfish, an open-source LB solver is built dynamically at run-time in CUDA or OpenCL \cite{januszewski2014sailfish}.
However, these two APIs require substantial changes in the original code, thus threatening the code’s correctness, portability, and maintainability.
The third accelerating approach is using Open Accelerators (OpenACC), which is a platform-independent, high level and directive-based programming standard. The programmers provide hints as annotations to the original code, specifying that a certain loop should run in parallel on the accelerator, then compute-intensive calculations are offloaded to the accelerator device without the need to explicitly manage data transfers between the host and the accelerator.
Recently, Open Multi-Processing (OpenMP) support for GPUs is also available.
Both OpenACC and OpenMP are directive based; the difference is that OpenACC is more descriptive, while OpenMP is more prescriptive (using distribute constructs to explicitly maps work-loads) \cite{calore2015accelerating}.

So far, there are fewer implementations of LB code using OpenACC compared to that using CUDA, primarily due to concerns regarding computational efficiency  \cite{calore2015accelerating,blair2015accelerating,calore2016performance}.
Xu et al. \cite{xu2017accelerated} demonstrated that for thermal flows (e.g., fluid flows and heat transfer in the side heated cavity), a speedup of around 53X can be achieved when using OpenACC acceleration compared with the serial code.
The results were obtained on a single Tesla K20c GPU, and 282.6 million lattice update per second (MLUPS, defined later in this paper) were reached for double-precision floating calculation.
Although higher performance could be gained using single-precision floating calculation \cite{kuznik2010lbm}, it would pose threats to the accuracy of the simulation results, particularly for turbulent flow simulations \cite{obrecht2013multi}.

To utilize the computing power of multi-node GPU clusters, LB implementations based on a hybrid OpenACC and Message Passing Interface (MPI) approach can be used for massively parallel simulations of problems with larger domain sizes.
However, simulations running on multiple GPUs have to face Peripheral Component Interconnect Express (PCI-e) bottlenecks and minimize inter-GPU communications.
In this work, we address implementation issues when using hybrid MPI and OpenACC to accelerate LB simulations.
We show that ultra-high computational performance can be achieved with proper implementations.
The rest of this paper is organized as follows.
In Section \ref{sec:numerical}, we introduce numerical details for the simulation of thermal convection, including the mathematical model and the corresponding LB model.
In Section \ref{sec:implementation}, we present details for implementation and optimization of the hybrid OpenACC and MPI, as well the parallel performance measured via the strong scaling test.
In Section \ref{sec:weak}, we present parallel performance measured via the weak scaling test.
In Section \ref{sec:conclusions}, the main findings of this work are summarized.

\section{Numerical method \label{sec:numerical}}
\subsection{Mathematical model for thermal convection}
We simulate thermal convection based on the Boussinesq approximation.
We assume the fluid flow is incompressible, and we treat the temperature as an active scalar that influences the velocity field through the buoyancy.
The viscous heat dissipation and compression work are neglected, and all the transport coefficients are assumed to be constants.
Then, the governing equations can be written as
\begin{subequations}
    \begin{gather}
        \nabla \cdot \mathbf{u}=0 \\
        \frac{\partial \mathbf{u}}{\partial t}+\mathbf{u} \cdot \nabla \mathbf{u}=-\frac{1}{\rho_0} \nabla P+\nu \nabla^2 \mathbf{u}+g \beta_T\left(T-T_0\right) \hat{\mathbf{z}} \\
        \frac{\partial T}{\partial t}+\mathbf{u} \cdot \nabla T=\alpha_T \nabla^2 T
    \end{gather}
    \label{eq:1}
\end{subequations}
where $\mathbf{u}$, $P$ and $T$ are the velocity, pressure, and temperature of the fluid, respectively.
$\rho_0$ and $T_0$ are reference density and temperature, respectively.
$\nu$, $\beta_T$ and $\alpha_T$ denote the viscosity, thermal expansion coefficient, and thermal diffusivity of the fluid, respectively.
$\hat{\mathbf{z}}$ is the unit parallel to gravity.
With the scaling
\begin{equation}
    \begin{aligned}
    &\mathbf{x}^*=\mathbf{x} / H, \quad t^*=t / \sqrt{H /\left(\beta_T g \Delta_T\right)}, \quad \mathbf{u}^*=\mathbf{u} / \sqrt{\beta_T g H \Delta_T}, \\
    &P^*=P /\left(\rho_0 g \beta_T \Delta_T H\right), \quad T^*=\left(T-T_0\right) / \Delta_T
    \end{aligned}
\end{equation}
Then, Eq. \ref{eq:1} can be rewritten in dimensionless form as
\begin{subequations}
    \begin{gather}
        \nabla \cdot \mathbf{u}^*=0 \\
        \frac{\partial \mathbf{u}^*}{\partial t^*}+\mathbf{u}^* \cdot \nabla \mathbf{u}^*=-\nabla P^*+\sqrt{\frac{\operatorname{Pr}}{R a}} \nabla^2 \mathbf{u}^*+T^* \tilde{\mathbf{z}} \\
        \frac{\partial T^*}{\partial t^*}+\mathbf{u}^* \cdot \nabla T^*=\sqrt{\frac{1}{\operatorname{PrRa}}} \nabla^2 T^*
    \end{gather}
\end{subequations}
Here, $H$ is the cell height and it is chosen as the characteristic length.
$t_f=\sqrt{H /\left(\beta_T g \Delta_T\right)}$ is the free-fall time and it is chosen as the characteristic time.
$\Delta_T$ is the temperature difference between heating and cooling walls.
The two dimensionless parameters are the $Ra$ and the $Pr$, which are defined as
\begin{equation}
    Ra =\frac{g \beta_T \Delta_T H^3}{\nu \alpha_T}, \ \  \ Pr=\frac{\nu}{\alpha_T}
\end{equation}

\subsection{The LB model for thermal convection}
\begin{figure}[!h]
    \centering
    \includegraphics[width = \textwidth]{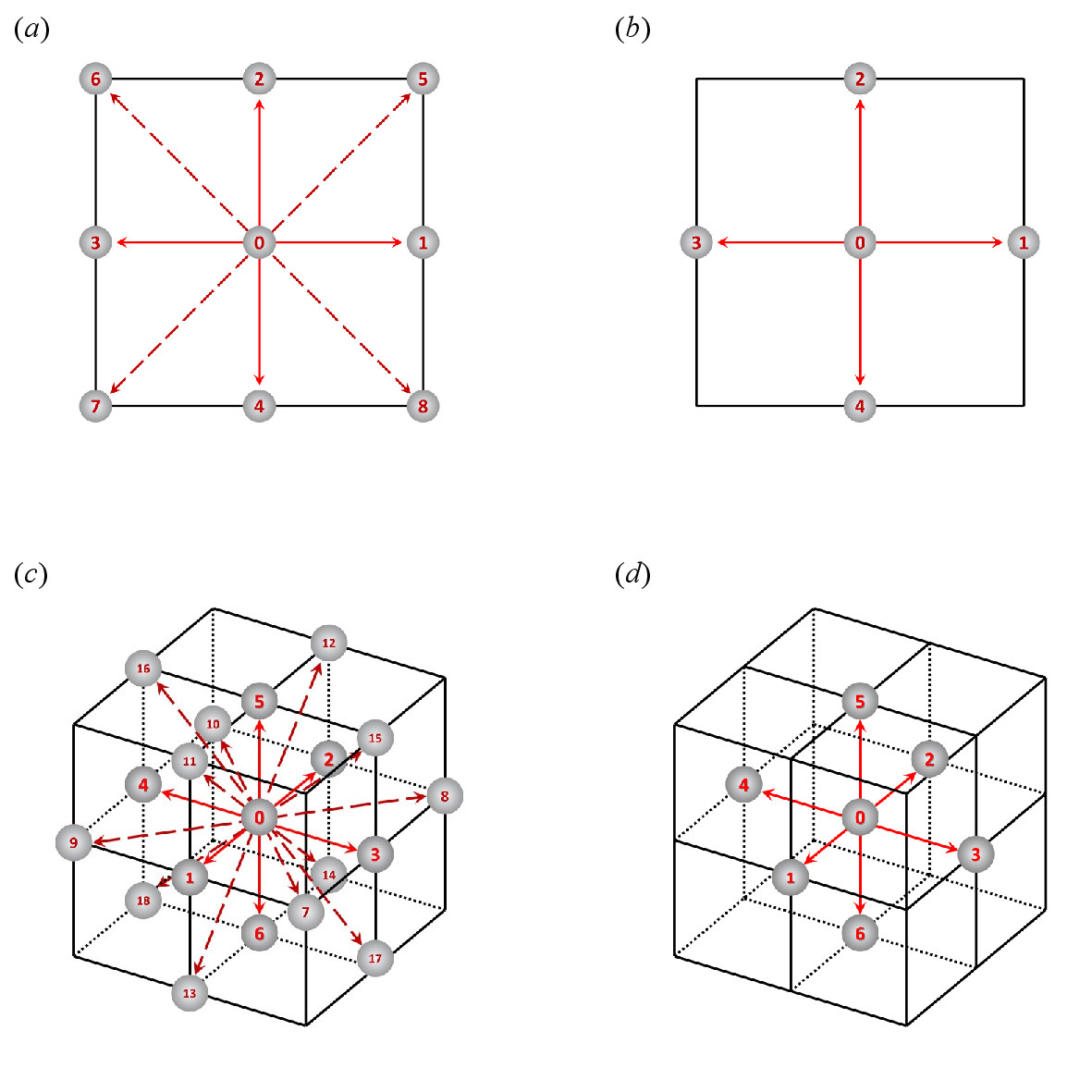}
    \caption{Illustration of discrete velocity models: (\textit{a}) the D2Q9 model and (\textit{b}) the D2Q5 model for two-dimensional simulations; (\textit{c}) the D3Q19 model and (\textit{d}) the D3Q7 model for three-dimensional simulation.}
    \label{fig:DnQm}
\end{figure}

We adopt the double distribution function (DDF)-based LB model to simulate thermal convective flows with the Boussinesq approximation \cite{yoshida2010multiple,chai2013lattice,wang2013lattice,contrino2014lattice}.
Specifically, we chose a D2Q9 discrete lattice in two-dimension (2D) or a D3Q19 discrete lattice in three-dimension (3D) for the Navier–Stokes equations to simulate fluid flows, and a D2Q5 discrete lattice in 2D or a D3Q7 discrete lattice in 3D for the energy equation to simulate heat transfer \cite{xu2017accelerated,xu2019lattice}, as illustrated in Fig. \ref{fig:DnQm}.
Here, the D2Q5 or D3Q7 model was chosen for the convection-diffusion equation to pursue computational efficiency.

To enhance the numerical stability, the multi-relaxation-time (MRT) collision operator is adopted in the evolution equations of both density and temperature distribution functions.
The evolution equation of the density distribution function is written as
\begin{equation}
f_i\left(\mathbf{x}+\mathbf{e}_i \delta_t, t+\delta_t\right)-f_i(\mathbf{x}, t)=-\left(\mathbf{M}^{-1} \mathbf{S}\right)_{i j}\left[\mathbf{m}_j(\mathbf{x}, t)-\mathbf{m}_j^{(\mathrm{eq})}(\mathbf{x}, t)\right]+\delta_t F_i^{\prime} \label{eq:f}
\end{equation}
where $f_i$ is the density distribution function  and $i=0,1,\cdots,q-1$.
$\mathbf{x}$ is the fluid plarcel position, $t$ is the time, $\delta_t$ is the time step. $\mathbf{e}_i$ is the discrete velocity along the $i$th direction.
For D2Q9 discrete lattice, $q=9$; for D3Q19 discrete lattice, $q=19$.
The forcing term $F_i^{\prime}$ on the right-hand side of Eq. \ref{eq:f} is given by $\mathbf{F}^{\prime}=\mathbf{M}^{-1}\left(\mathbf{I}-\frac{\mathbf{S}}{2}\right) \mathbf{M \tilde{F}}$, and the term $\mathbf{M \tilde { F }}$ is given as \cite{guo2002discrete,guo2008analysis,chai2012effect}
\begin{subequations}
    \begin{equation}
        \mathbf{M \tilde{F}}_{D 2 Q 9}=\left[0,6 \mathbf{u} \cdot \mathbf{F},-6 \mathbf{u} \cdot \mathbf{F}, F_x,-F_x, F_y,-F_y, 2 u F_x-2 v F_y, u F_x+v F_y\right]^T
    \end{equation}
    \begin{align}
        \begin{split}
            \mathbf{M \tilde{F} }_{ D 3 Q 1 9 } = & {\left[0,38 \mathbf{u} \cdot \mathbf{F},-11 \mathbf{u} \cdot \mathbf{F}, F_x,-\frac{2}{3} F_x, F_y,-\frac{2}{3} F_y, F_z,-\frac{2}{3} F_z,\right.} \\
        & 4 u F_x-2 v F_y-2 w F_z,-2 u F_x+v F_y+w F_z, 2 v F_y-2 w F_z, \\
        &\left.-v F_y+w F_z, u F_y+v F_x, v F_z+w F_y, u F_z+w F_x, 0,0,0\right]^T
        \end{split}
    \end{align}
\end{subequations}
where $\mathbf{F}=\rho g \beta_T\left(T-T_0\right) \hat{\mathbf{y}}$ in 2D or $\mathbf{F}=\rho g \beta_T\left(T-T_0\right) \hat{\mathbf{z}}$ in 3D.
The macroscopic density $\rho$ and velocity $\mathbf{u}$ are obtained from $\rho=\sum_{i=0}^{q-1} f_i, \mathbf{u}=\left(\sum_{i=0}^{q-1} \mathbf{e}_i f_i+\frac{1}{2} \mathbf{F}\right)/\rho$.
At the fluid-solid boundary, the no-slip velocity boundary condition can be realized via the half-way bounce-back scheme $f_{\bar{i}}\left(\mathbf{x}, t+\delta_t\right)=f_i^{+}(\mathbf{x}, t)$.
Here, $f_i^{+}(\mathbf{x}, t)$ denotes the post-collision density distribution function; $f_{\bar{i}}(\mathbf{x}, t)$ denotes the density distribution function associated with the velocity $\mathbf{e}_{\bar{i}}$, and we have the relation $\mathbf{e}_{\bar{i}}=-\mathbf{e}_i$.

The evolution equation of the temperature distribution function is written as
\begin{equation}
g_i\left(\mathbf{x}+\mathbf{e}_i \delta_t, t+\delta_t\right)-g_i(\mathbf{x}, t)=-\left(\mathbf{N}^{-1} \mathbf{Q}\right)_{i j}\left[\mathbf{n}_j(\mathbf{x}, t)-\mathbf{n}_j^{(\mathrm{eq})}(\mathbf{x}, t)\right] \label{eq:g}
\end{equation}
where $g_i$ is the temperature distribution function and $i=0,1,\cdots,q-1$.
For D2Q5 discrete lattice, $q=5$; for D3Q7 discrete lattice, $q=7$.
The macroscopic temperature $T$ is obtained from $T=\sum_{i=0}^{q-1} g_i$.
At the fluid-solid boundary, the Dirichlet temperature boundary condition can be realized via the half-way anti-bounce-back scheme $g_{\bar{i}}\left(\mathbf{x}, t+\delta_t\right)=-g_i^{+}(\mathbf{x}, t)+\omega T_w$, where $T_w$ is the
wall temperature, $\omega=\left(4+a_T\right) / 10$ for D2Q5 and $\omega=\left(6+a_T\right) / 21$ for D3Q7, $a_T$ is a constant related to thermal diffusivity \cite{li2013boundary};
the Neumann adiabatic boundary condition can be realized via the half-way bounce-back scheme $g_{\bar{i}}\left(\mathbf{x}, t+\delta_t\right)=g_i^{+}(\mathbf{x}, t)$.
Here, $g_i^{+}(\mathbf{x}, t)$ denotes the post-collision temperature distribution function; $g_{\bar{i}}(\mathbf{x}, t)$ denotes the temperature distribution function associated with the velocity $\mathbf{e}_{\bar{i}}$.
More numerical details of the thermal LB method can be found in our previous work \cite{xu2017accelerated,xu2019lattice}.

\subsection{Simulation settings and parallel performance characterization}
\begin{figure}[!h]
    \centering
    \includegraphics[width = \textwidth]{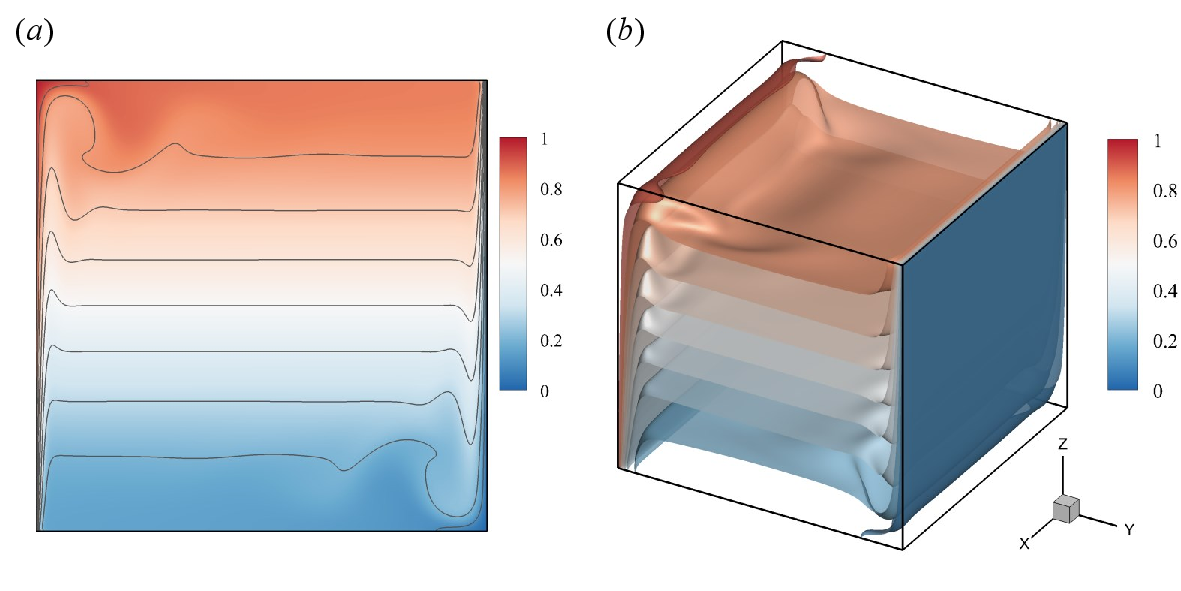}
    \caption{The temperature field in the convection cell: (\textit{a}) the 2D simulation and (\textit{b}) the 3D simulation.}
    \label{fig:temperatureField}
\end{figure}

We consider a 2D square cell and a 3D cubic cell.
The left and right walls of the cell are kept at constant hot and cold temperatures, respectively; while the other two (or four) walls of the 2D cell (or the 3D cell) are adiabatic; all walls impose no-slip velocity boundary conditions.
We fixed the Prandtl number as $Pr = 0.71$.
For the 2D thermal convection, the Rayleigh number is fixed as $Ra = 10^8$; while for the 3D thermal convection, the Rayleigh number is fixed as
$Ra = 10^7$.
Previously, we obtained a steady flow pattern in these two cases \cite{xu2017accelerated,xu2019lattice}; while at higher $Ra$, the flow transits to an unsteady state, and a Hopf bifurcation occurs.
We verified the multi-GPU implementation can give correct simulation results consistent with our previous results \cite{xu2017accelerated,xu2019lattice} (see Fig. \ref{fig:temperatureField} for the temperature field in the convection cell); however, for the sake of clarity, we do not repeat to provide the tabulated flow quantities and heat transfer properties.
In the following, we focus on the parallel performance of the multi-GPU simulation.

In the strong scaling test, we measure the running time as a function of GPU numbers for fixed total problem size, which represents the ability to solve a problem faster using more resources;
while in the weak scaling test, we measure the running time as a function of GPU numbers for fixed problem size per GPU (i.e., the total problem size increases), which represents the ability to solve larger problems with larger resources.
We adopt two metrics to characterize the parallel performance of the LB simulation, one is Million Lattice Updates Per Second (MLUPS) and the other is parallel efficiency.
The MLUPS is defined as \cite{bailey2009accelerating}
\begin{equation}
    \text { MLUPS }=\frac{\text { mesh size } \times \text { iteration steps }}{\text { running time } \times 10^6}
\end{equation}
Meanwhile, we have 1 GLUPS = 1000 MLUPS, where GLUPS stands for billion lattice updates per second.
In the strong scaling test, the parallel efficiency $\eta$ is defined as
\begin{equation}
\eta=\frac{T_1}{T_n \cdot n}
\end{equation}
Here, $T_1$ denotes the running time using a single GPU and $T_n$ denotes the running time using $n$ GPUs.
In the weak scaling test, the parallel efficiency $\eta$ is simply calculated as $\eta=T_1 / T_n$.
We conducted experiments on a four-node GPU cluster, each node is equipped with four NVIDIA A100 GPUs powered by the NVIDIA Ampere Architecture.
The network interconnects use 100 Gigabits per second (Gbps) Remote direct memory access over Converged Ethernet (RoCE).
The inter-GPU-GPU communication within a node goes over the PCI-e.
To automatically utilize the GPUDirect acceleration technologies, we adopt a CUDA-aware MPI implemented in OpenMPI.
With GPUDirect technology, including Peer to Peer (P2P) and Remote Direct Memory Access (RDMA), the buffers can be directly sent from a GPU memory to another GPU memory or the network without touching the host memory \cite{ye2022accelerating}.

We measure the running time and then calculate the MLUPS and parallel efficiency.
In Section \ref{sec:implementation}, we describe optimization strategies for multi-GPU simulation and evaluate the parallel performance via the strong scaling test;
then, in Section \ref{sec:weak}, we further discuss the parallel performance via the weak scaling test.
In the strong scaling test, we fix a grid number of $8193\times 8193$ in the 2D simulation
and a grid number of $385\times 385\times 385$ in the 3D simulation.
Here, we use an odd number of grid points in each dimension for two reasons: first, it reduces the oscillations due to spurious conserved quantities \cite{luo2011numerics};
secondly, for a grid number of $N$ in each dimension, it allows the $[(N+1)/2]^\text{th}$ point to be precisely located at the center line (plane) based on the half-way (anti-)bounce-back boundary scheme, which is convenient for post-analysis of flow and heat transfer quantities.
The iteration steps are fixed as 12,000 and 6000, respectively, for the 2D and 3D simulations.
Preliminary tests showed that the corresponding running time was around 30 seconds with 16 GPUs (i.e., the largest number of GPUs in the tests), which ensures the measurement of computing performance enters a steady state.
Detailed settings for the weak scaling test are described in Section \ref{sec:conclusions}.
All the simulations adopt double-precision floating-point arithmetic, which ensures the simulation accuracy.

\section{Implementation and optimization of the hybrid OpenACC and MPI approach \label{sec:implementation}}

\subsection{Naive implementation of hybrid OpenACC and MPI approach}

\begin{figure}[!h]
    \centering
    \includegraphics[width = \textwidth]{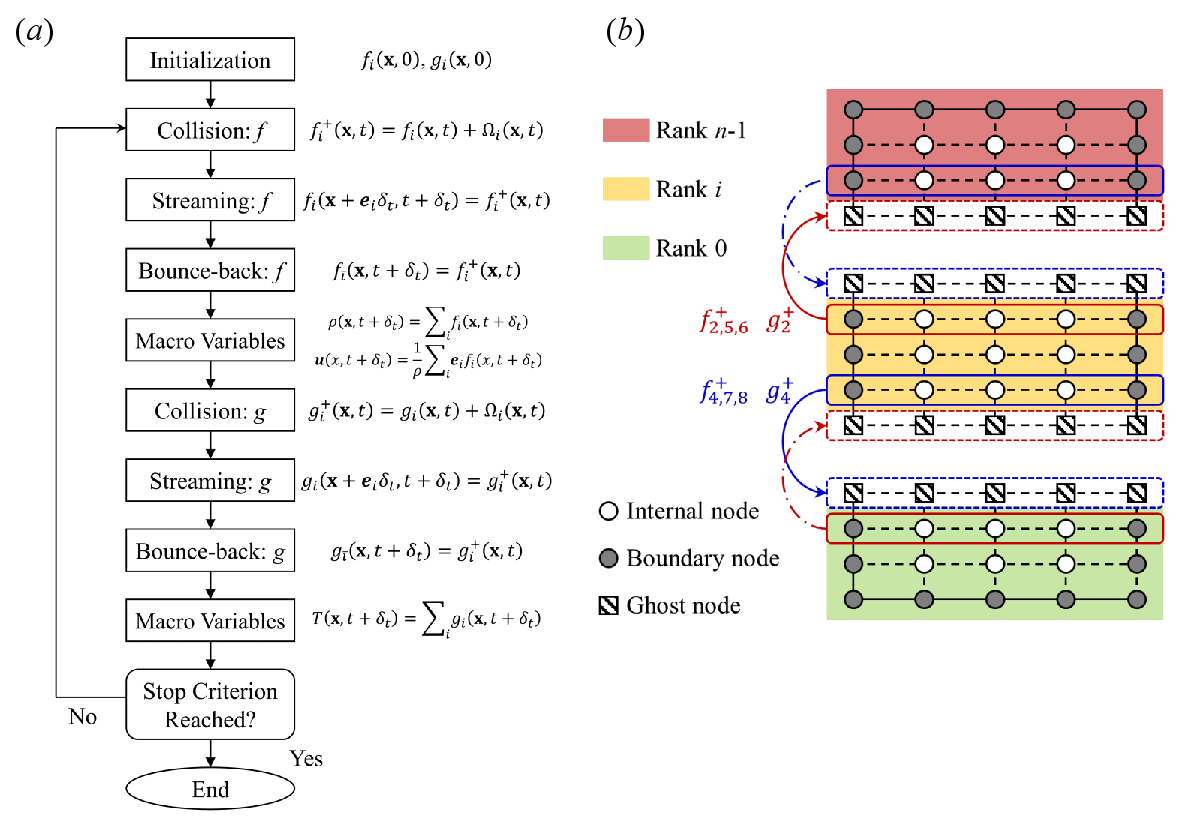}
    \caption{(\textit{a}) Flowchart of double distribution function (DDF)-based lattice Boltzmann (LB) model for thermal convection simulation;
    (\textit{b}) mono-dimensional partitioning of the 2D computation domain and the associated data exchange.}
    \label{fig:mono_illustration}
\end{figure}

To utilize the computing power of GPU clusters, we use the hybrid OpenACC and MPI approach, in which OpenACC accelerates the computation on a single GPU and MPI synchronizes the information between multiple GPUs.
In Fig. \ref{fig:mono_illustration}(a), we present the flowchart of the DDF-based LB model for thermal convection problems.
Here, we split the evaluation of the density distribution function (i.e., Eq. \ref{eq:f}) into the following two steps
\begin{subequations}
    \begin{align}
        &Collision \  step: f_i^{+}(\mathbf{x}, t)=f_i(\mathbf{x}, t)-\left(\mathbf{M}^{-1} \mathbf{S}\right)_{i j}\left[\mathbf{m}_j(\mathbf{x}, t)-\mathbf{m}_j^{(\mathrm{eq})}(\mathbf{x}, t)\right]+\delta_t F_i^{\prime}   \\
        &Streaming \ step: f_i\left(\mathbf{x}+\mathbf{e}_i \delta_t, t+\delta_t\right)=f_i^{+}(\mathbf{x}, t)
    \end{align}
\end{subequations}
Similarly, we split the evaluation of the temperature distribution function (i.e., Eq. \ref{eq:g}) into two steps as
\begin{subequations}
    \begin{align}
        &Collision \  step: g_i^{+}(\mathbf{x}, t)=g_i(\mathbf{x}, t)-\left(\mathbf{N}^{-1} \mathbf{Q}\right)_{i j}\left[\mathbf{n}_j(\mathbf{x}, t)-\mathbf{n}_j^{(\mathrm{eq})}(\mathbf{x}, t)\right]   \\
        &Streaming \  step: \mathrm{g}_i\left(\mathbf{x}+\mathbf{e}_i \delta_t, t+\delta_t\right)=g_i^{+}(\mathbf{x}, t)
    \end{align}
\end{subequations}
In the collision step, we can use the symbol $\mathbf{\Omega}_i(\mathbf{x},t)$ to denote the collision operator, which is the second term on the right-hand side of the equation.
The collision step is also known as the relaxation step and it completely updates local information;
the streaming step is also known as the propagation step and it transfers data information to its neighboring.
Because GPUs adopt single instruction multiple threads (SIMT) execution model, we optimized the data layout for the distribution function and stored the information in the structure of array (SoA) to meet the requirement of coalescing memory access, such that neighboring threads access neighboring data.
For example, the density distribution function $f_i(\mathbf{x}, t)$ is stored with index $(x + N_{x} \times y + N_{x} \times N_{y} \times i)$ or $(x + N_{x} \times y + N_{x} \times N_{y} \times z + N_{x} \times N_{y} \times N_{z} \times i)$, respectively, for the 2D or 3D case.
Here, $N_{x}$, $N_{y}$, and $N_{z}$ denote the grid number in the $x$, $y$, and $z$ directions, respectively.
Using the column-major order programming language (such as Fortran), the corresponding index formula implementation is $f(x, y, i)$ or  $f(x, y, z, i)$  \cite{xu2017accelerated}.

In a naive implementation of the hybrid OpenACC and MPI approach, we adopt a mono-dimensional partitioning of the computational domain.
Take the column-major order programming language (such as Fortran) as an example, we decompose the domain along the $y$-direction (or the $z$-direction) in the 2D (or the 3D) domain, then the interface between the subdomains is along the $x$-direction line (or the $x$-$y$ plane), which favors the continuous transfer data in memory space.
We add ghost layers outside the boundary of each subdomain to receive data from the adjacent subdomains, as illustrated in Fig. \ref{fig:mono_illustration}(b).
To reduce the amount of data transfer, in the streaming step, we do not transfer the full array of the distribution function $f_i^{+}(\mathbf{x})$ or $g_i^{+}(\mathbf{x})$, but only components of these arrays corresponding to specific discrete velocity directions.
For example, using the D2Q9 discrete lattice for the density distribution function and D2Q5 discrete lattice for the temperature distribution function, the current GPU [the corresponding subdomain is marked by orange color in Fig. \ref{fig:mono_illustration}(b)] only sends values of $f_{2,5,6}^{+}(\mathbf{x})$ and $g_2^{+}(\mathbf{x})$ [these are distribution functions belong to the top boundary of the subdomain, see the red rectangle with the solid line in Fig. \ref{fig:mono_illustration}(b)] to the top neighbor GPU,
while it only sends values of $f_{4,7,8}^{+}(\mathbf{x})$ and $g_4^{+}(\mathbf{x})$ [these are distribution functions belong to the bottom boundary of the subdomain,
see the blue rectangle with the solid line in Fig. \ref{fig:mono_illustration}(b)] to the bottom neighbor GPU.

\begin{figure}[!h]
    \centering
    \includegraphics[width = \textwidth]{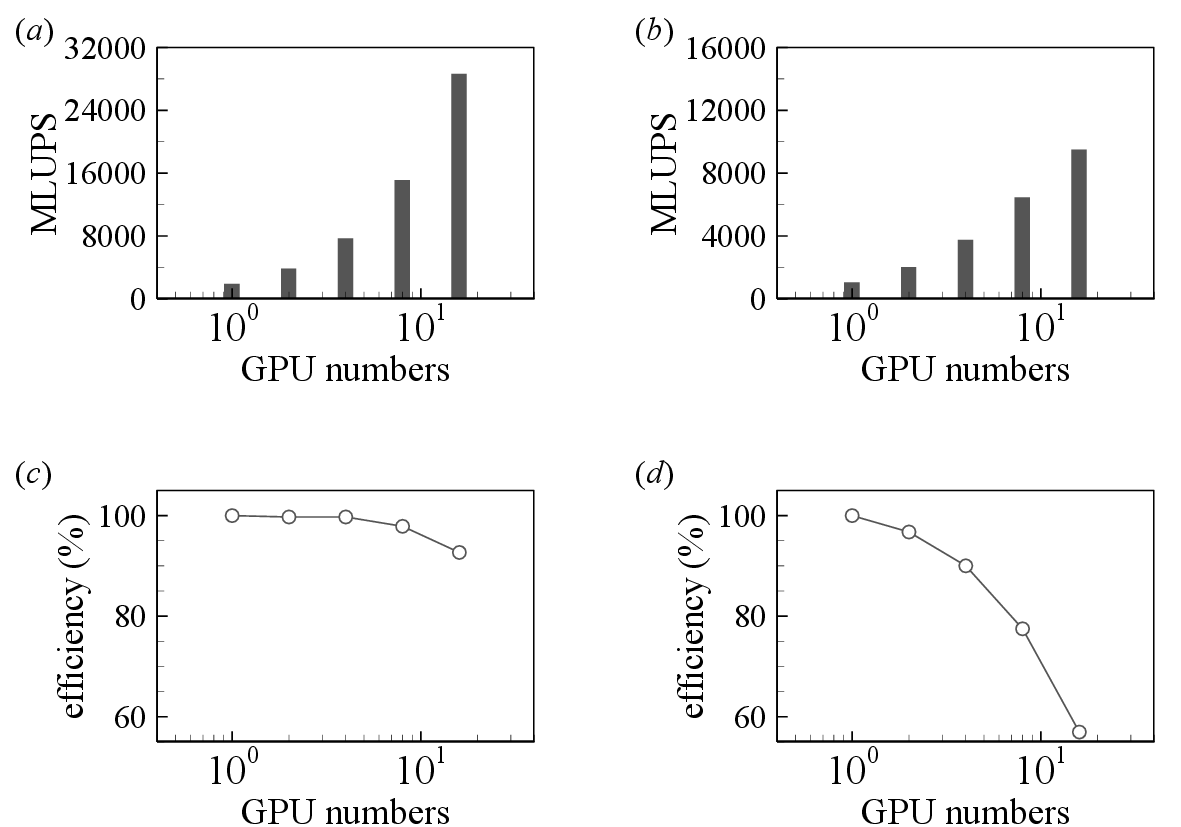}
    \caption{Performance of (\textit{a}, \textit{c}) the 2D simulation and (\textit{b}, \textit{d}) the 3D simulation, in terms of (\textit{a}, \textit{b}) the MLUPS and (\textit{c}, \textit{d}) the parallel efficiency.}
    \label{fig:mono}
\end{figure}

Figure \ref{fig:mono} shows the parallel performance in terms of MULPS and parallel efficiency for the 2D simulation and the 3D simulation, respectively.
We can see that using 1 GPU, the 2D thermal LB simulation achieves 1931.2 MLUPS, and the 3D thermal LB simulation achieves 1042.4 MLUPS.
We note in the D2Q9 + D2Q5 thermal LB model, each grid node accesses 80 variables within an iteration step and these variables occupy 8 bytes in double-precision;
while in the D3Q19 + D3Q7 thermal LB model, each grid node accesses 143 variables within an iteration step.
Because the NVIDIA A100 has a memory bandwidth of 1555 GB/s, the thermal LB simulation has a theoretical maximum performance of $1555 \times 1000^3 / (80 \times 8 \times 10^6) = 2429.7$ MLUPS for the 2D simulation, and $1555 \times 1000^3 / (143 \times 8 \times 10^6) = 1359.3$ MLUPS for the 3D simulation.
In other words, we have reached 79.5\% and 76.7\% of the theoretical maximum performance, similar to that of the SunwayLB \cite{liu2019sunwaylb}.
With the increase of GPU numbers, the MULPS generally increases;
however, the parallel efficiency degrades when more GPUs are used, particularly for the 3D case in which the parallel efficiency is less than 57\% using 16 GPUs, indicating the parallel code is not scalable.
Two reasons may be responsible for the poor parallel performance. First, in the strong scaling measurement, the computational load on each GPU decreases with the increase of GPU numbers; however, the communication cost per GPU almost remains the same.
Adopting the mono-dimensional partitioning, the amount of data that needs to be transferred is $2 \times N_{x}$ and $2 \times N_{x} \times N_{y}$, respectively, for the 2D and the 3D simulation.
As a result, the ratio between communication time and the total time increases with the increase of GPU numbers, restricting the scalability of the code.
Secondly, if the grid number of the whole computation domain is not large enough, the computational load on each GPU is not sufficient to fully occupy the resources of that GPU, and the overhead to launch kernel matters.
To further boost the parallel performance using multi-GPUs, we describe some optimization strategies in the following subsections.

\subsection{Block partitioning of the computational domain}

\begin{figure}[!h]
    \centering
    \includegraphics[width = 0.7\textwidth]{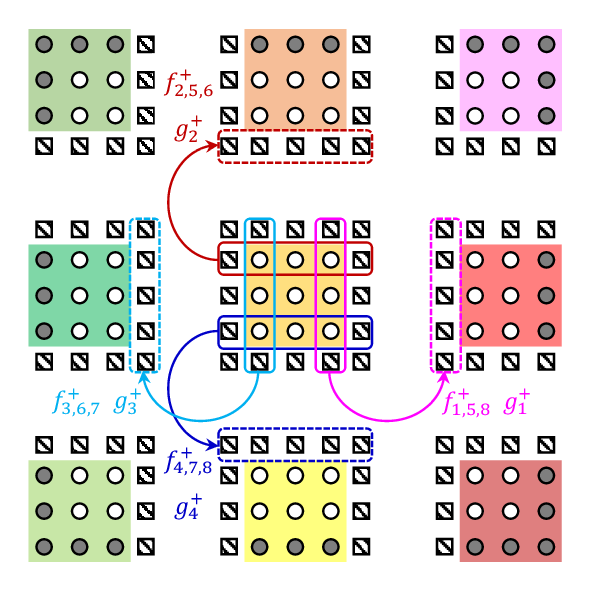}
    \caption{Illustration of block partitioning and data exchange for the 2D domain.}
    \label{fig:block_illustration}
\end{figure}

In mono-dimensional partitioning, the computational domain is decomposed into several slices and each slice is allocated to a single GPU.
An alternative strategy for domain decomposition is block partitioning, which is to decompose the domain in more than one dimension \cite{schepke2009parallel,obrecht2013scalable}.
Figure \ref{fig:block_illustration} illustrates the block partitioning and data exchange for the 2D domain.
We decompose the domain both along the $x$ and the $y$ directions, and we minimize the differences in the subdomain size along each dimension.
Using the column-major order programming language (such as Fortran), the data is stored continuously along the $x$-direction in the memory space, we preferably decompose the domain along the $y$-direction (in 2D) or the $z$-direction (in 3D).
The partition details for the domain in 2D and 3D are provided in Table \ref{tb:partition}.
After that, we add ghost layers to each subdomain to store the data received from adjacent subdomains.
Using Fortran, we need to wrap the left and right boundary nodes along the $y$-direction into a contiguous cache space before sending them to its left and right neighboring GPUs (see the nodes in the light blue and purple rectangles with solid lines in Fig. \ref{fig:block_illustration}).
Here we also included ghost nodes outside the subdomain boundaries when sending the message, such that data for the corner points can be implicitly synchronized and transferred to the diagonally neighboring GPUs.
With CUDA-aware MPI libraries implemented in OpenMPI, we discourage the use of MPI-derived datatypes for data communication, even though both contiguous and non-contiguous derived datatypes are supported.
The non-contiguous datatypes currently have a high overhead because of the many calls to copy all the pieces of the buffer into the intermediate buffer.
Previously, Calore et al. \cite{calore2016massively} overcame this issue by developing a custom communication library that uses persistent send and receives buffers, allocated once on the GPUs at program initialization.
Here, we recommend an alternative solution that simply put the ghost nodes into a buffer of contiguous memory locations.

\begin{table}
\centering
\caption{Partition details of the domain in 2D and 3D.}
\begin{tabular}{ccc}
  \hline
  GPU numbers & Partition type in 2D & partition type in 3D \\
  \hline
  1 & - & - \\
  2 & $1\times 2$ & $1\times 1 \times 2$ \\
  4 & $2\times 2$ & $1\times 2 \times 2$ \\
  8 & $2\times 4$ & $2\times 2 \times 2$ \\
  16 & $4\times 4$ & $2\times 2 \times 4$ \\
  \hline
\end{tabular} \label{tb:partition}
\end{table}

\begin{figure}[!h]
    \centering
    \includegraphics[width = \textwidth]{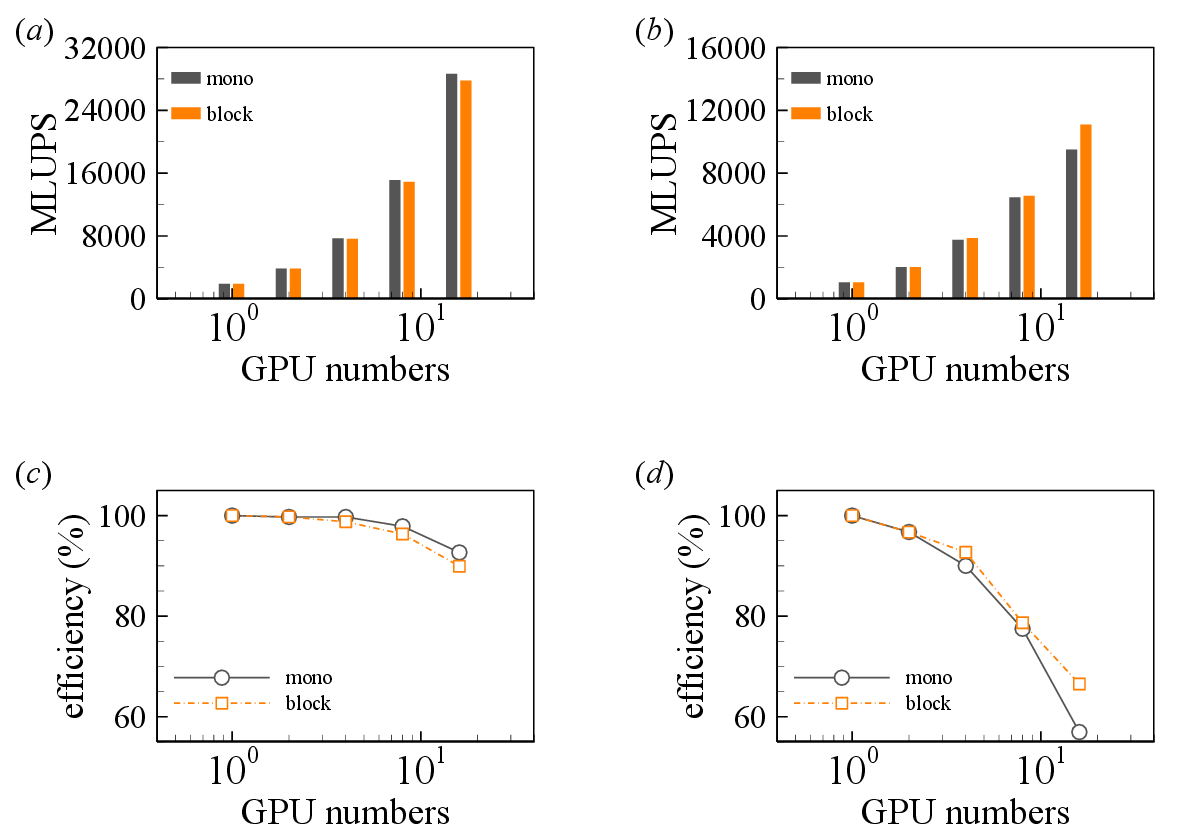}
    \caption{Performance comparisons between mono-dimensional and block partitioning of the computational domain for  (\textit{a}, \textit{c}) the 2D simulation and (\textit{b}, \textit{d}) the 3D simulation, in terms of (\textit{a}, \textit{b}) the MLUPS and (\textit{c}, \textit{d}) the parallel efficiency.}
    \label{fig:block}
\end{figure}

Figure \ref{fig:block} compares the performance between mono-dimensional partitioning and block partitioning of the computational domain.
We can see that adopting block partitioning can improve the MLUPS and parallel efficiency for the 3D simulations, however, it slightly degrades the parallel performance for the 2D simulations.
In the following, we provide a theoretical analysis of the advantages and disadvantages of adopting block partitioning.
For the 2D computational domain with size $N_{x} \times N_{y}$, we assume a square domain of $N_{x} = N_{y} = N$ for simplicity.
Adopting the mono-dimensional partitioning, each GPU sends 2$N$ boundary nodes to its neighboring GPUs, and it launches two kernels for data communications on the two boundaries;
while adopting the block partitioning, each GPU sends $4(N/\sqrt{a}+2)$ boundary nodes to its neighboring, and it launches four kernels for data commutations on the four boundaries.
Here, for simplicity, we discuss the case when the GPU number $a$ meets the requirement that $\sqrt{a}$ is an integer.
We denote that the time consumption for sending a single boundary node is $k$, and the time consumption to launch a kernel for data communication is $h$.
Thus, for mono-dimensional and block partitioning, the total time consumption for message passing, which includes the time for sending data and the time to launch a kernel, on a single GPU is $t_{1} = 2kN+2h$ and $t_{2} = 4(N/\sqrt{a}+2)k+4h$, respectively.
The above analysis indicates that the block partitioning would be more efficient when $t_{1} > t_{2}$, i.e., $2kN-4Nk/\sqrt{a} > 2h+8k$;
in other words, we prefer to use block partitioning when both the domain size $N$ and the GPU number $a$ are large.

\subsection{Overlapping communications with computations\label{sec:overlapping}}

\begin{figure}[!h]
    \centering
    \includegraphics[width = \textwidth]{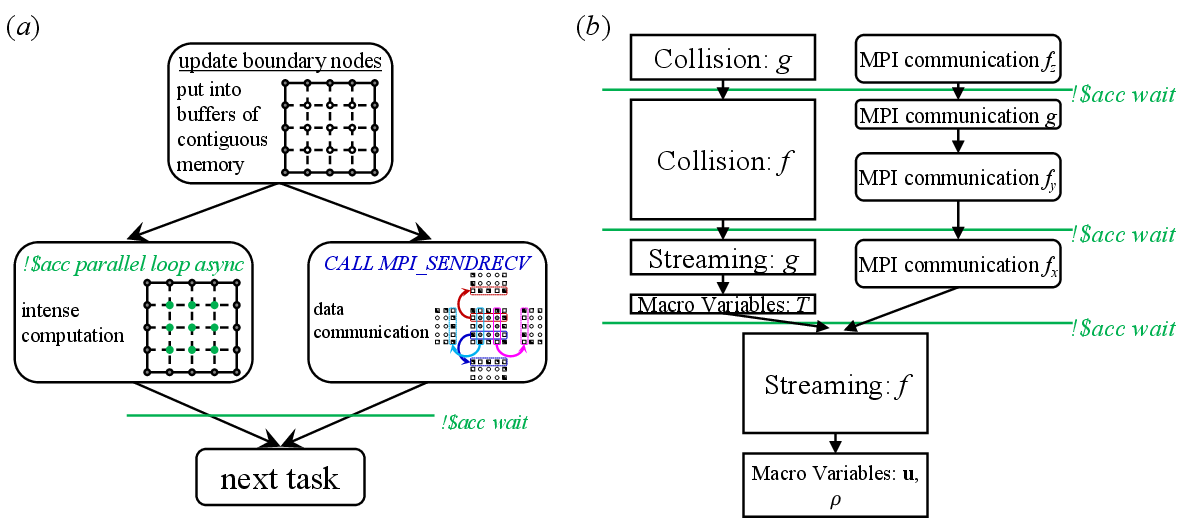}
    \caption{Schematic illustration of (\textit{a}) the basic idea to overlap communications with computations in OpenACC and (\textit{b}) an example in the thermal LB model for overlapping communications with computations.
    The size of the box to represent each computation subroutine is generally proportional to its time consumption (obtained via NVIDIA Nsight Systems).
    The bounce-back routines updating the distribution functions at boundaries are neglected because they account for less than 1\% of the total time.}
    \label{fig:overlap_illustration}
\end{figure}

Minimizing the communication overhead is crucial to improve the parallel performance of the simulations based on multi-GPUs, we further hide communication overhead behind the kernel runtime by overlapping the communications with computations.
The basic idea is to simultaneously execute kernels on GPUs for intense computation and data communications among GPUs, because GPUs cannot control data transfer and only CPUs can manage the data communication.
On devices with distinct hosts and device memory, we can create asynchronous work queues to deal with the cost of PCIe data transfers.
Practically, we can adopt the  \emph{!\$acc async} handle in OpenACC.
As illustrated in Fig. \ref{fig:overlap_illustration}(a), we first update the boundary nodes and put them into buffers of contiguous memory.
We then create asynchronous work queues to perform intense computation for updating inner nodes, while synchronizing data of boundary nodes among various GPUs using MPI.
Thus, the latency of communication can be hidden behind the computations.
The next task will not begin execution until all actions on the async queues are complete, which can be realized using the \emph{!\$acc wait} handle in OpenACC.
In Fig. \ref{fig:overlap_illustration}(b), we give an example of the thermal LB model for overlapping communications with computations.
Because the evolution of density distribution function $f_{i}$ and temperature distribution function $g_{i}$ are independent of each other within an iteration, we can update $f_{i}$ and $g_{i}$ alternatively to hide the communication overhead.
Meanwhile, the amount of data computation for $f_{i}$ is larger than that for $g_{i}$, since there are 5 components of $f_{i}$ for the D3Q19 discrete lattice and 1 component of $g_{i}$ for the D3Q7 discrete lattice on a surface of the subdomain, we then divide the communication of $f_{i}$ into three sets: $f_{x}$ representing the density distribution at the back and front surfaces, $f_{y}$ representing the density distribution function at the left and right surfaces, and $f_{z}$ representing the density distribution function at the bottom and top surfaces.
As illustrated in Fig. \ref{fig:overlap_illustration}(b), we first relax $g_{i}$ at the inner nodes while communicating $f_{z}$.
After that, we communicate $g_{i}$ at the boundary nodes while relaxing $f_{i}$ at the inner nodes.
Because the relaxation of $f_{i}$ is the most time-consuming subroutine, as evident by the results obtained via NVIDIA Nsight Systems and represented by the box size in the illustration drawing, we can also hide the communication of $f_{y}$ while relaxing $f_{i}$ at the inner nodes.
For the communication of $f_{x}$, it can be hidden by the propagation step of $g_{i}$ and the step to calculate macroscopic variables of $T$.
Now that the data communication for $f_{i}$ and $g_{i}$ at the boundary nodes are completed within an iteration step, we do not have to overlap communications for the remaining computation subroutines of propagation $f_{i}$ and calculation macroscopic variables of $\mathbf{u}$ and  $\rho$.

\begin{figure}[!h]
    \centering
    \includegraphics[width = \textwidth]{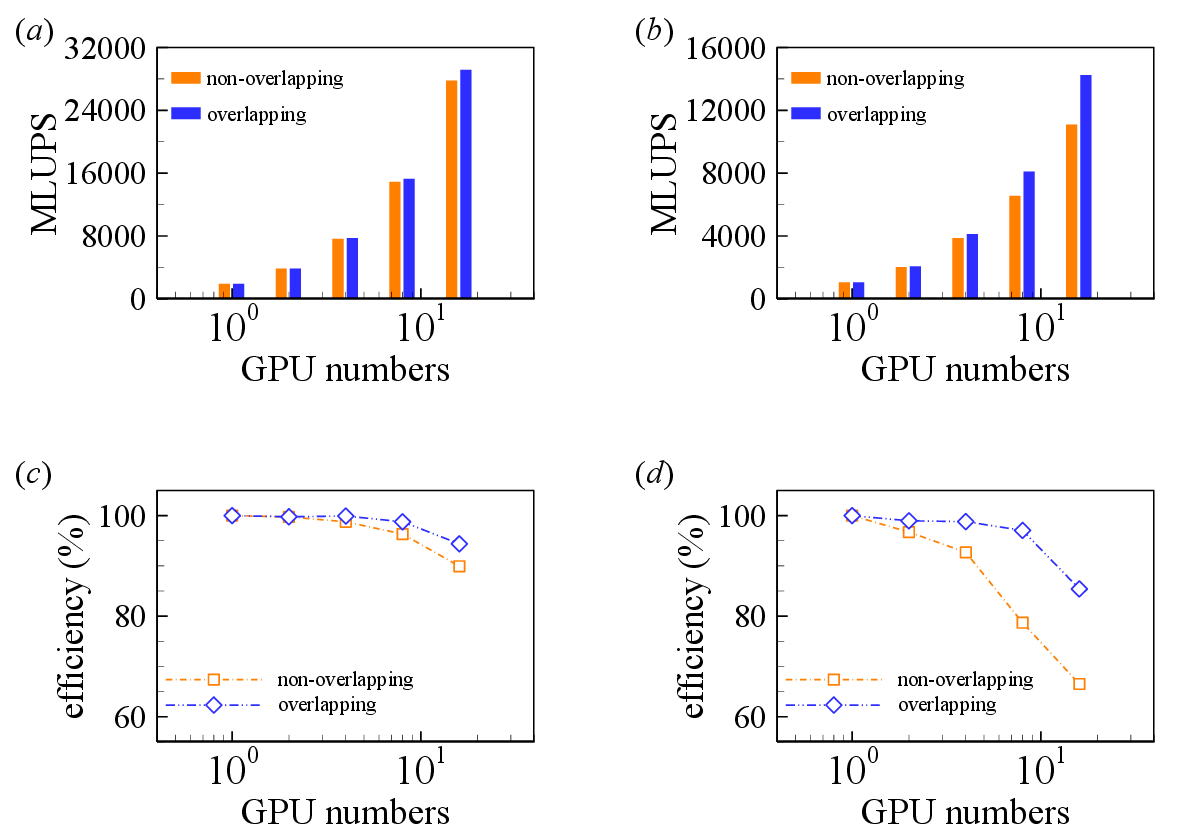}
    \caption{Performance comparisons between non-overlapping and overlapping communication with computation for (\textit{a}, \textit{c}) the 2D simulation and (\textit{b}, \textit{d}) the 3D simulation, in terms of (\textit{a}, \textit{b}) the MLUPS and (\textit{c}, \textit{d}) the parallel efficiency.}
    \label{fig:overlap}
\end{figure}

Figure \ref{fig:overlap} compares the performance of non-overlapping and overlapping communication with computation.
We can see that the MLUPS and parallel efficiency improved for all the cases if the communications and computations overlapped, and the advantage of using the overlapping mode is more obvious with the increase of GPU numbers.
For the 3D simulations, the performance improvement is greater than that in the 2D simulations, because more time for data transfer between GPU communication is hidden.
Excitingly, for the 3D simulation, the parallel efficiency improves from 78.7\% to 97.1\% when using 8 GPUs, and the MLUPS increases by 1.28X when using 16 GPUs.
Previously, using the overlapping mode implemented in a hybrid CUDA and MPI approach, Xian and Takayuki \cite{xian2011multi} reported the MLUPS increased by 1.24X with 16 GPUs,
Hong et al. \cite{hong2015scalable} reported the MLUPS increased by 1.38X with 6 GPUs, suggesting a similar amount of increase in the parallel performance even though we adopt the high level and directive-based OpenACC standard.

\subsection{Concurrent computation on a GPU}
\begin{figure}[!h]
    \centering
    \includegraphics[width = \textwidth]{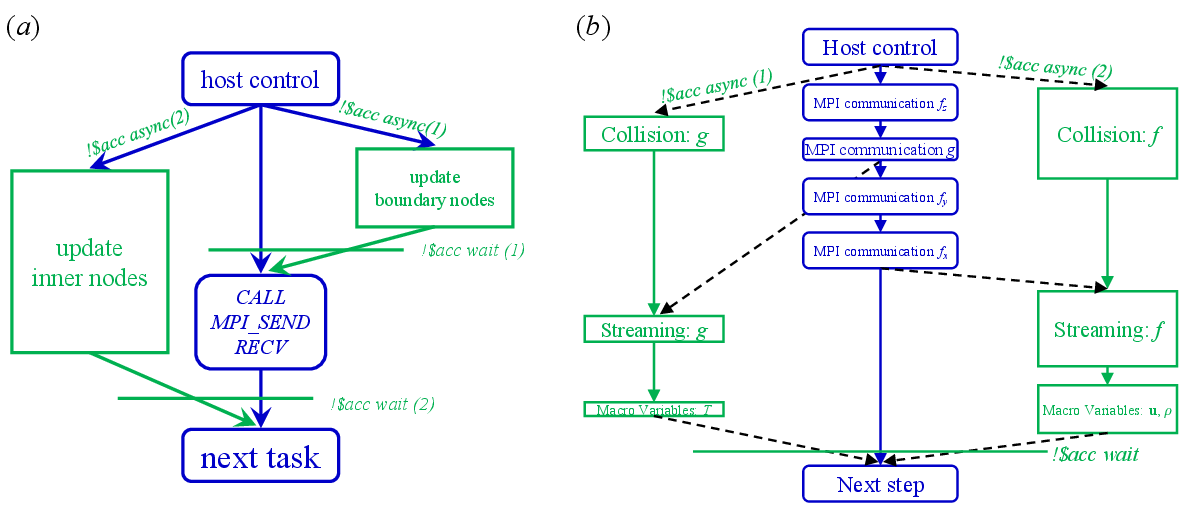}
    \caption{Schematic illustration of (\textit{a}) the basic idea of concurrently executing two independent tasks in OpenACC and (\textit{b}) an example in the thermal LB model for concurrent computation.}
    \label{fig:concurrent_illustration}
\end{figure}

In addition to data parallelism, exploiting task parallelism can further utilize the GPU hardware resources, particularly when using a large number of GPUs, the computational tasks may not fully occupy the resources of the GPU, and we can concurrently execute two independent tasks on a single GPU, as illustrated in Fig. \ref{fig:concurrent_illustration}(a).
Practically, we can adopt the \emph{!\$acc async (n)} handle, which launches work asynchronously in queue $n$.
All operations in the same queue will execute in order, while operations in different queues may execute in any order.
In Fig. \ref{fig:concurrent_illustration}(b), we give an example of the thermal LB model for concurrent computation.
Because the evolution of density distribution function $f_{i}$ and temperature distribution function $g_{i}$ are independent of each other within an iteration, we can also update $f_{i}$ and $g_{i}$ in different queues.
It should be noted that the data communication at the boundary nodes should be performed before the propagation step.
Here, following the discussion in Section \ref{sec:overlapping}, we divide the communication of $f_{i}$ into three sets, and we overlap the communication of $f_{i}$ and $g_{i}$ [see the blue box in Fig. \ref{fig:concurrent_illustration}(b)] with computation [see the green box in Fig. \ref{fig:concurrent_illustration}(b)].

\begin{figure}[!h]
    \centering
    \includegraphics[width = \textwidth]{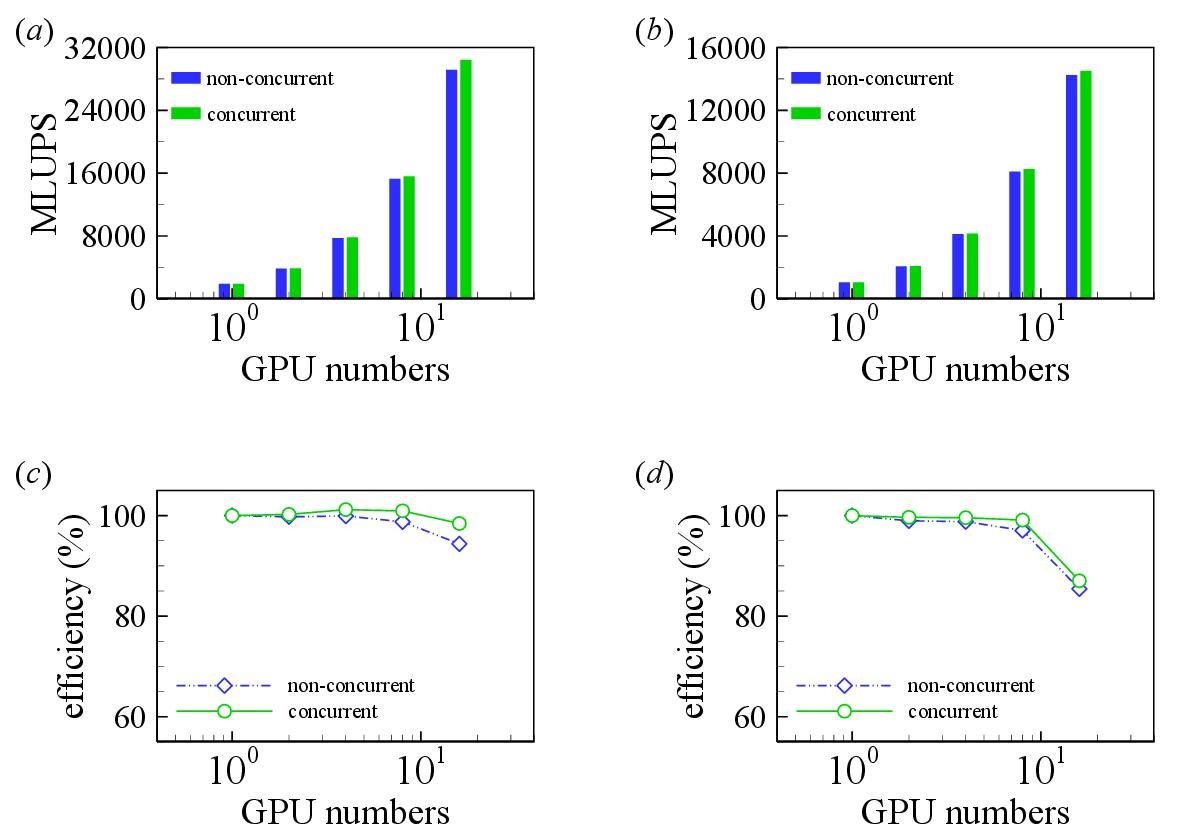}
    \caption{Performance comparisons between non-concurrent and concurrent computation for (\textit{a}, \textit{c}) the 2D simulation and (\textit{b}, \textit{d}) the 3D simulation, in terms of (\textit{a}, \textit{b}) the MLUPS and (\textit{c}, \textit{d}) the parallel efficiency.}
    \label{fig:concurrent}
\end{figure}

Figure \ref{fig:concurrent} compares the performance of non-concurrent and concurrent computation.
We can see the MLUPS and the parallel efficiency slightly improves.
Using 16 GPUs, the 2D simulation achieved 30.42 GLUPS, and the 3D simulation achieved 14.52 GLUPS.
An interesting finding is that for the 2D simulation, the parallel efficiency may even be greater than 100\%, which may be due to the efficient use of the on-chip memory on the GPU, highlighting the advantage of exploiting task parallelism.
Due to the high ability of GPU for computation, the GPU performance for the 3D simulation can be further boosted with the increasing of computational load.

\section{Weak scaling test\label{sec:weak}}
\begin{figure}[!h]
    \centering
    \includegraphics[width = \textwidth]{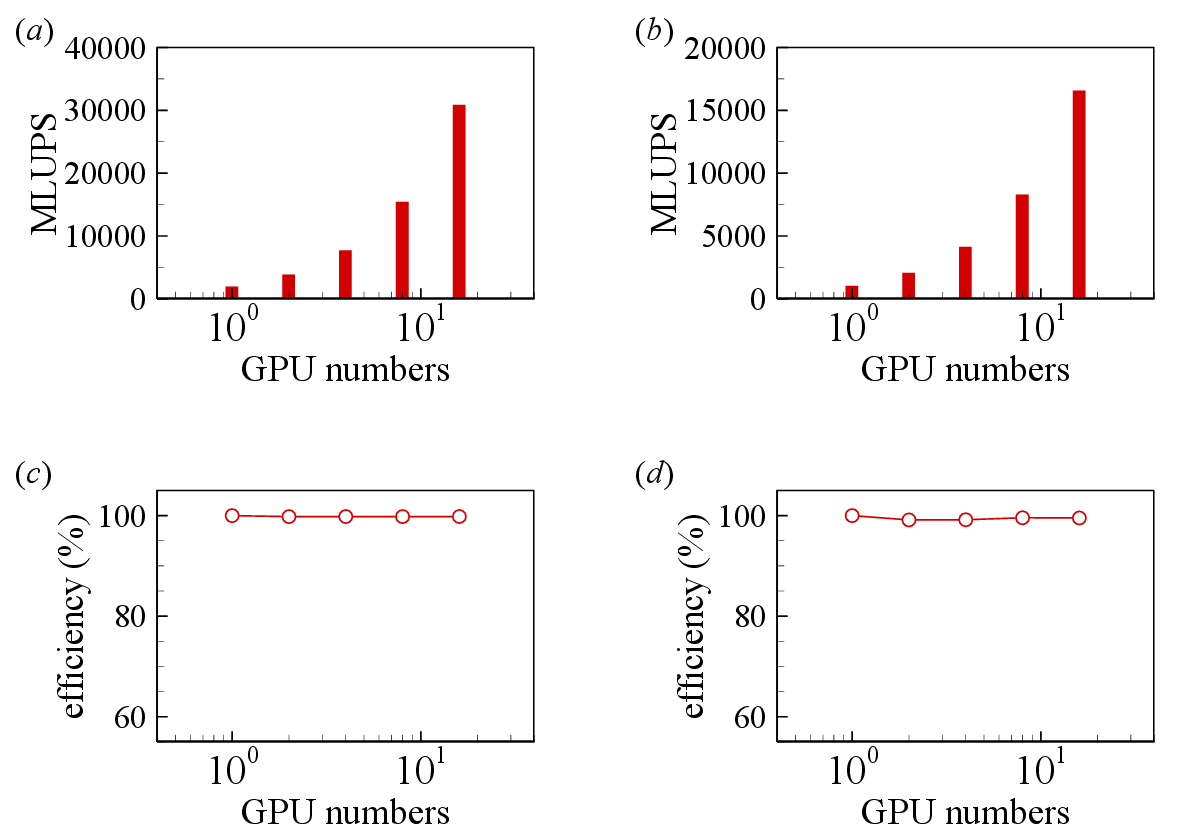}
    \caption{Weak scaling test: performance of (\textit{a}, \textit{c}) the 2D simulation and (\textit{b}, \textit{d}) the 3D simulation, in terms of (\textit{a}, \textit{b}) the MLUPS and (\textit{c}, \textit{d}) the parallel efficiency.}
    \label{fig:weak_scaling}
\end{figure}

In the weak scaling test, we use a sub-domain size of $8192\times 8192$ and $384\times 384\times 384$ in the 2D and the 3D simulation, respectively; the iteration steps are fixed as 1000 (in 2D) and 500 (in 3D).
Each dimension of the domain size increases similarly to that of increasing in GPU numbers (see Table \ref{tb:partition} for the partition details of the domain).
For simplicity, we only provide the performance of the code after adopting all the optimization strategies described in Section \ref{sec:implementation}.
As shown in Fig. \ref{fig:weak_scaling}, using 1 GPU, we can achieve 1932.9 MLUPS and 1045.3 MLUPS in the 2D and 3D simulation (less than 0.3\% deviation from that in the strong scaling test due to different runs), respectively.
With the increase of GPU numbers, the MLUPS almost increases linearly up to 16 GPUs, and the parallel efficiency remains above 99\%. These results demonstrate that the optimized thermal LB code has excellent weak scalability.

\section{Conclusions\label{sec:conclusions}}

In this work, we have adopted a hybrid OpenACC and MPI approach for accelerated thermal LB simulation on multi-GPUs.
The OpenACC accelerates computation on a single GPU, and the MPI synchronizes the information between multiple GPUs.
We adopt the double distribution function-based thermal LB model, namely, D2Q9 + D2Q5 in the 2D simulation, and D3Q19 + D3Q7 in the 3D simulation.
With a single NVIDIA A100 GPU, the 2D simulation achieved 1.93 GLUPS with a grid number of $8193^{2}$ and the 3D thermal LB simulation achieves 1.04 GLUPS with a grid number of $385^{3}$, which is more than 76\% of the theoretical maximum performance.
In a naive implementation to extend to multi-GPUs, we used mono-direction partitioning of the computation domain, however, the code was not scalable to more than 8 GPUs in the 3D simulation.
To further boost the parallel performance, we adopted three optimization strategies: block partitioning, overlapping communications with computations, and concurrent computation.

With block partitioning, the domain is decomposed in more than one dimension, and it decreases the amount of data that needs to be transferred.
By overlapping the communications with computations, communication overhead is hidden behind the kernel runtime.
Using concurrent computation, task parallelism can be exploited to better utilize the GPU hardware resources.
After adopting these optimization strategies, we demonstrate that the parallel performance can be significantly improved.
In the strong scaling test, using 16 GPUs, the 2D simulation achieved 30.42 GLUPS and the 3D simulation achieved 14.52 GLUPS.
In the weak scaling test, the parallel efficiency remains above 99\% up to 16 GPUs.
It should be noted that all performance measurements are based on double-precision floating-point arithmetic, which ensures simulation accuracy.
Our results demonstrated that, with improved data and task management, the hybrid OpenACC and MPI technique is promising for thermal LB simulation on multi-GPUs.

\section*{Appendix A. Comparison of the parallel performance on CPUs versus GPUs}
\begin{figure}[!h]
    \centering
    \includegraphics[width = \textwidth]{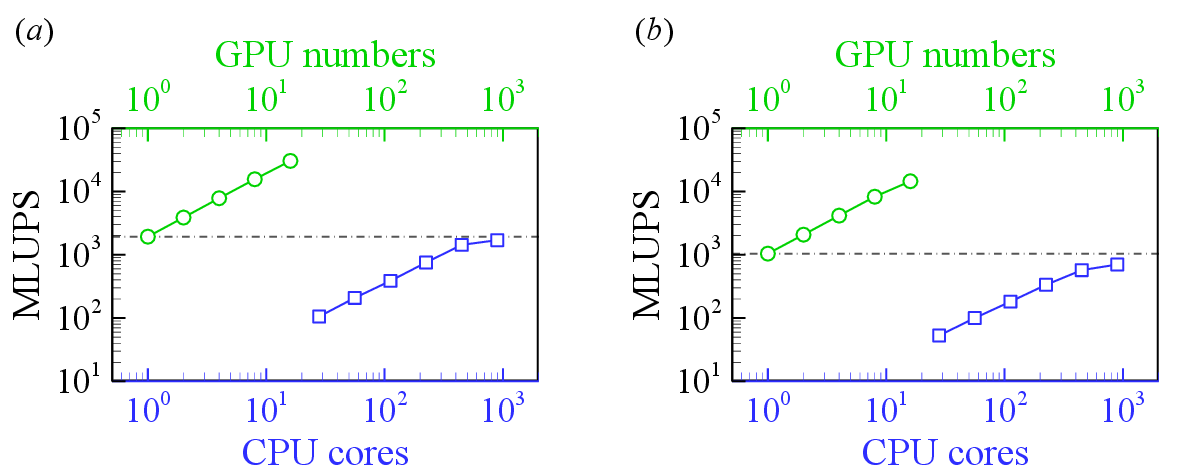}
    \caption{Comparison of the parallel performance on CPUs versus GPUs for (\textit{a}) the 2D simulation and (\textit{b}) the 3D simulation in terms of the MLUPS.
    The gray-dashed lines denote the MLUPS on a single GPU.
    The total number of CPU cores is $N_{1} \times N_{2}$, where $N_{1}$ is the number of MPI processes and $N_{2}$ is the number of OpenMP threads per MPI process.
    Here, we fix $N_{2}$ as 14 and $N_{1}$ increases from 2 to 64.}
    \label{fig:CPUvsGPU}
\end{figure}

In the appendix, we summarize the MLUPS of the GPU code adopting the optimization strategies discussed in Section \ref{sec:implementation}.
As a comparison, we provide the MLUPS of a CPU code, which is developed with a hybrid MPI and OpenMP approach \cite{jin2011high}.
The CPU code is based on a hierarchical two-level parallelization, where the first-level parallelization applies MPI domain decomposition to the simulation domain, and the second-level parallelization uses OpenMP parallel regions for loops within a subdomain.
Such a hybrid MPI + OpenMP approach can reduce the memory usage and overhead associated with MPI calls.
For experiments on the CPU cluster, each node is equipped with two Intel Xeon 6258R CPUs (i.e., 56 cores within a node), we assign 4 MPI processes on each node with 14 OpenMP threads per MPI process.
This choice is a compromise between two factors: first, more OpenMP threads reduces the corresponding number of MPI processes, which leads to better communication performances; secondly, more OpenMP threads increases the overhead associated with OpenMP constructs and remote memory accesses across sockets.
For all the cases investigated, our test results showed that on each node, the combination of 4 MPI processes $\times$ 14 OpenMP threads works slightly better than that of 7 MPI processes $\times$ 8 OpenMP threads.
We can see from Fig. \ref{fig:CPUvsGPU} that in the strong scaling test, both our CPU code and GPU code scale well.

\section*{Acknowledgements}
This work was supported by the National Natural Science Foundation of China (NSFC) through Grant Nos. 11902268 and 12272311,
the Open Fund of Key Laboratory of Icing and Anti/De-icing (Grant No. IADL20200301).
and the National Key Project via No. GJXM92579,
The authors acknowledge the Beijing Beilong Super Cloud Computing Co., Ltd for providing HPC resources that have contributed to the research results reported within this paper (URL: http://www.blsc.cn/).

\section*{References}


\end{document}